# Giant magnetocaloric effect in $Co_2FeAl$ Heusler alloy nanoparticles

**Aquil Ahmad[*1], Srimanta Mitra[1,2], S. K. Srivastava[1], and A. K. Das[*1]**

[1]*Department of Physics, Indian Institute of Technology Kharagpur, Kharagpur, India-721302*

[2]*Space Applications Center, ISRO, Ahmedabad, India-380015*

[*]*Corresponding Authors*

*E-mails: aquil@phy.iitkgp.ernet.in and amal@phy.iitkgp.ernet.in*

## Abstract

A giant magnetocaloric effect across the ferromagnetic (FM) to paramagnetic (PM) phase transition was observed in chemically synthesized $Co_2FeAl$ Heusler alloy nanoparticles with a mean diameter of 16 nm. In our previous report, we have observed a significant enhancement in its saturation magnetization ($M_s$) and Curie temperature ($T_c$) as compared with the bulk counterpart. Motivated from those results, here, we aim to explore its magnetocaloric properties near the $T_c$. The magnetic entropy change ($-\Delta S_M$) shows a positive anomaly at 1252 K. $-\Delta S_M$ increases linearly with the magnetic field, and a large value of ~15 J/Kg-K is detected under a moderate field of 14 kOe. It leads to a net relative cooling power of 89 J/Kg for the magnetic field change of 14 kOe. To confirm the nature of magnetic phase transition, a detailed study of its magnetization is performed. The Arrott plot and nature of universal curve conclude that FM to PM phase transition in the present system is of second-order.

## 1. Introduction

Heusler alloys are the materials of enormous interest due to their multifunctional properties viz. magnetoresistance (MR) [1], half-metallicity (i.e., 100 spin polarization at the Fermi edge) [2-7], spin filtering [8], spin injection [9], shape memory [10, 11] and thermoelectric effects [12]. These Half-metallic ferromagnets (HMF) exhibit a unique band structure at the Fermi energy ($E_F$), where one spin channel shows metallic nature and another spin channel displays a non-conducting or semiconducting gap. An interesting character of Heusler alloys is that the structural, magnetic and magnetocaloric properties can easily be tuned by controlling their elemental compositions or partial substitution by other elements [13-16]. In 1983, de Groot *et al.*, firstly reported such behavior in NiMnSb Heusler alloy [17]. The $X_2YZ$ (X/Y: transition metals; Z: main group element) Heusler compounds, generally known as full-Heusler alloy, may be crystalized in the cubic $L2_1$ (i.e., fully-ordered) structures under the space group $Fm\bar{3}m$ (#225) [18]. However, the ideal $L2_1$ phase is difficult to achieve, especially in nanoscale regime; therefore, they are generally crystalized in either partial-disordered (i.e., B2 type) [19] or fully disordered (i.e., A2 type) [20] phases. The half-metallic nature, large magnetic moment and high $T_c$ of Heusler compounds (specially, in $Co_2$-based), make them potential candidates for spintronic applications [20-22]. Besides, the magnetocaloric (MC) effect has also been explored in Heusler alloys (HAs) [23-25]. The materials that can potentially be used as magnetic refrigerants near to or above room temperature are of great interest [13, 26, 27]. In this regard, materials exhibiting giant magnetocaloric effect (GMCE) would be promising [16, 24, 28-30]. The GMCE was initially observed in $Gd_5(Si_2Ge_2)$ [29, 31]. Nevertheless, the high cost of Gadolinium makes it unsuitable for commercial magnetic refrigerator. Moreover, there are many other key challenges from the materials perspective to improve their performance and implementation in devices [32]. The peak of the entropy change ($-\Delta S_M$) was found to be significantly higher in HAs [24, 33, 34] with first-order magnetic phase transition, which is necessary for efficient cooling. However, thermal and magnetic hysteresis present in such systems limits their practical usefulness. This further motivates researchers to search for other Heusler based materials. Lately, few works on MC properties of Co-based HAs have been reported [13, 35, 36]. They exhibit second-order phase transition and are found to be usable as magnetic refrigerants due to their broad working temperature range and absence of thermal and magnetic hysteresis. The spin orbit coupling (SOC) is stronger in Co-based HAs [37]; therefore, the magnetocrystalline anisotropic exchange interaction might be responsible in their magnetic ordering. Practically, there is no system available, which strictly follows the Stoner's itinerant model or Heisenberg theory of the localized spin. No particular theory can completely explain the magnetism of such 3d intermetallic compounds. Therefore, it is necessary to explore their magnetism near the phase transition.

As a young research field, earlier studies on Heusler nanoalloys have been focused on their synthesis, structure, and magnetic characterizations (see refs. [20, 38], and the references there in). Consequently, further investigations are required to verify their technical importance in magnetic refrigeration and microactuators. Previously, we have reported size modulated structural and magnetic properties of $Co_2FeAl$ (CFA) nanoparticles (NPs). CFA-NPs crystallized in cubic A2-disorder phase. Their microstructural analysis reveals that NPs are spherical and crystalline in nature with an average particle size of 16 nm. Moreover, they exhibit pronounced enhancement in $M_s$ and $T_c$ as compared with their bulk counterpart. A record magnetic moment and $T_c$ of 6.5 $\mu_B$/f.u. (25% higher than the bulk) and 1261 K (15% higher than the bulk) respectively, have been observed [20]. As a thumb rule, to get huge MC response, the material should have a higher magnetic moment [27]; therefore, a large magnetic entropy change ($-\Delta S_M$) is expected in CFA-NPs. Here, we report the MC effect of CFA-NPs, near the FM-PM phase transition.

## 2. Experimental details

CFA-NPs were synthesized using co-precipitation and thermal deoxidization method. The precursor salts were $CoCl_2{\cdot}6H_2O$ (99%), $Fe(NO_3)_3{\cdot}9H_2O$ (99%) and $Al_2(NO_3){\cdot}18H_2O$ (98%). The detailed description of the sample preparation was described in our previous report [20]. NPs were analyzed by x-ray diffraction (XRD), high-resolution transmission electron microscopy (HRTEM) and selected area electron diffraction (SAED) techniques. MC properties were studied using high-temperature vibrating sample magnetometer (VSM) of Lake Shore Cryotronics. In magnetization measurements, powder sample was placed into a disposable boron nitride (BN) cup as provided by the Lake Shore Cryotronics.

## 3. Results and discussion

The simulated and experimental x-ray diffraction data of CFA-NPs are shown in figure 1. The prominent Bragg peaks of (220), (400) and (422) establish that the present system is crystallized in A2 disordered phase, but not in $L2_1$ phase [20]. Note that (111) and (200) superlattice reflections, which are absent in our case, are required to conclude the $L2_1$ structure [7]. The lattice constant, as calculated from the noticeable (220) Bragg peak is 5.72 Å. This closely matches with the bulk value [39].

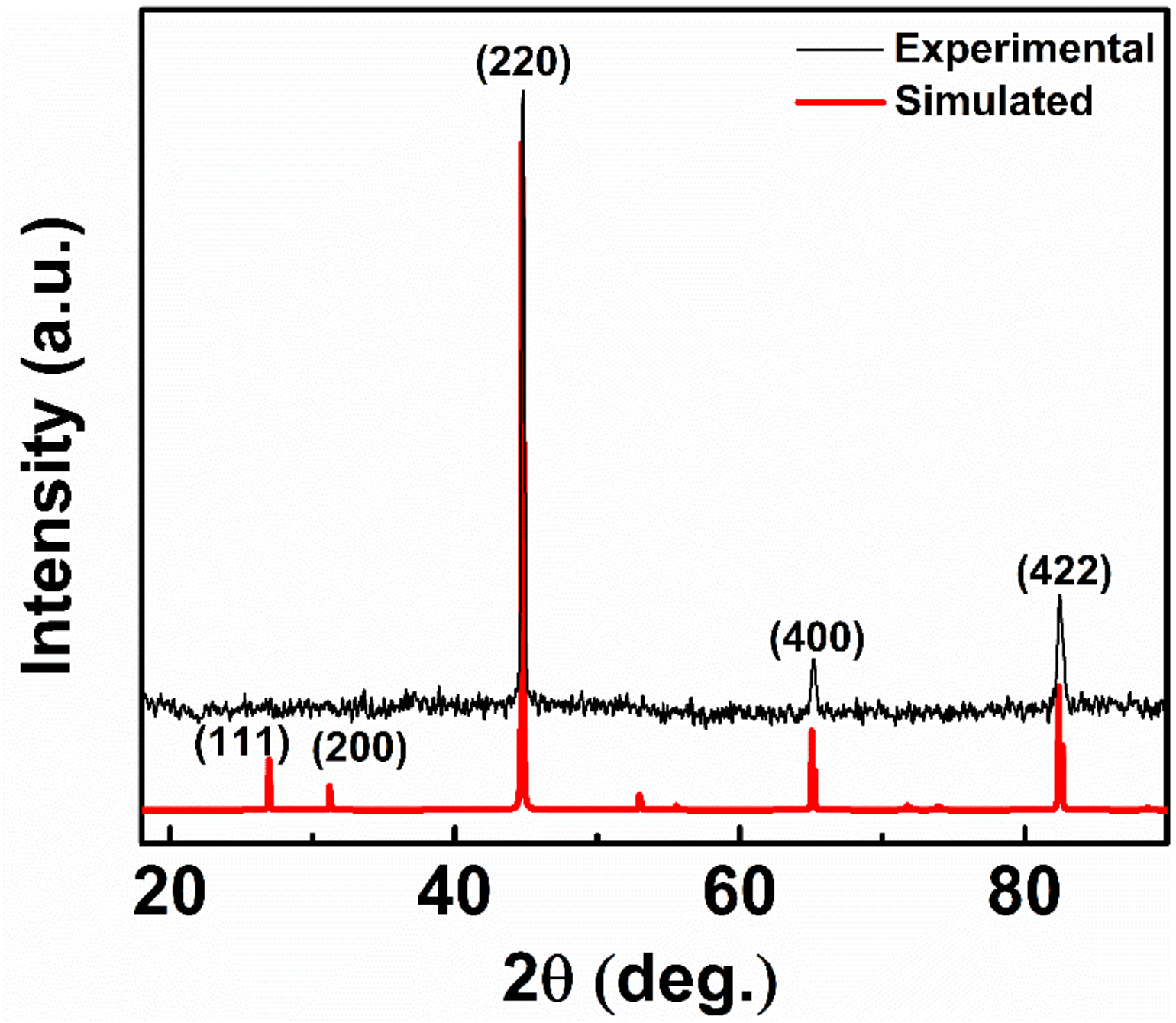

**Figure 1.** Experimental and simulated XRD patterns of $Co_2FeAl$ nanoparticles.

The NPs are analyzed using HRTEM imaging (figure 2). As evident from figure 2(a), particles are nearly spherical in shape. The image is further examined for particle size distribution by Image J software. The size-distribution histogram, along with the fitted Gaussian profile, is shown in figure 2 (b). According to the histogram, the NPs are of sizes 16 ± 10 nm. The SAED pattern (see figure 2 (c)) features concentric rings accompanied by dots. This indicates that particles are crystalline in nature. The image of a single nanoparticle with an even higher resolution is shown in figure 2(d). The presence of observable lattice fringes confirm the high crystallinity of the particles. An enlarged portion of the image has been presented in the inset of figure 2(d). The interplanar spacing of 2.022 Å is equivalent to the (220) plane of the cubic phase of CFA-NPs, and is consistent with the XRD results.

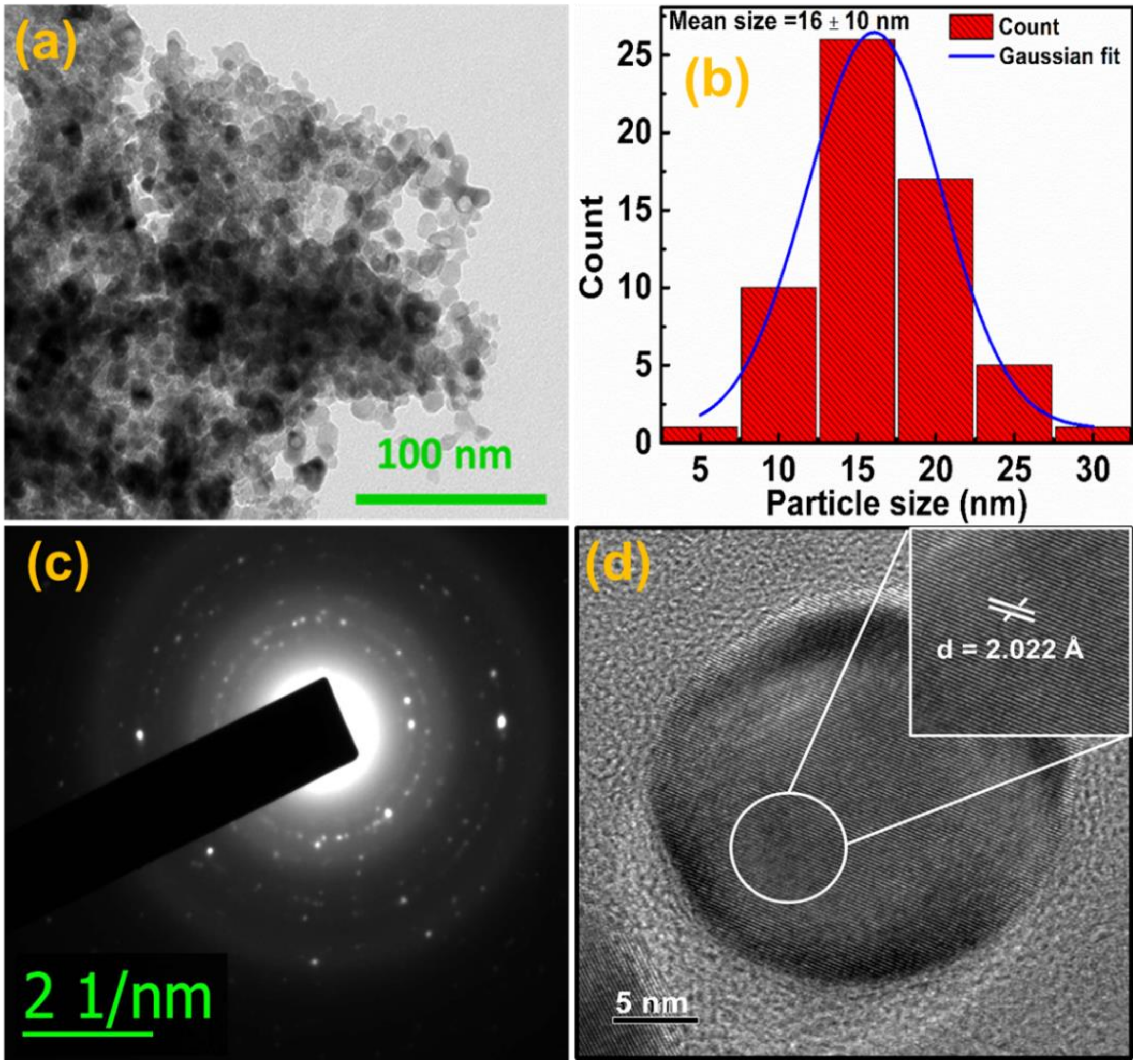

**Figure 2.** (a) Transmission electron microscopy image of CFA-NPs. (b) A histogram for the size distribution of NPs. (c) SAED pattern and (d) High-resolution images displaying the lattice planes and crystallinity; inset represents enlarged portion of the image.

Thermomagnetic zero-field-cooled (ZFC) and field-cooled (FC) curves of CFA-NPs measured at 500 Oe field, in the temperature range of 300 K-1273 K, are shown in figure 3. Both the ZFC and FC curves of CFA are smooth and exhibit a clear FM to PM phase transition around the $T_c$. The M(T) curves resemble a continuous or second order FM-PM phase transition [40]; this argument is further supported by the absence of the thermal hysteresis in the present system. This motivated us to further investigate its phase transition and MC properties.

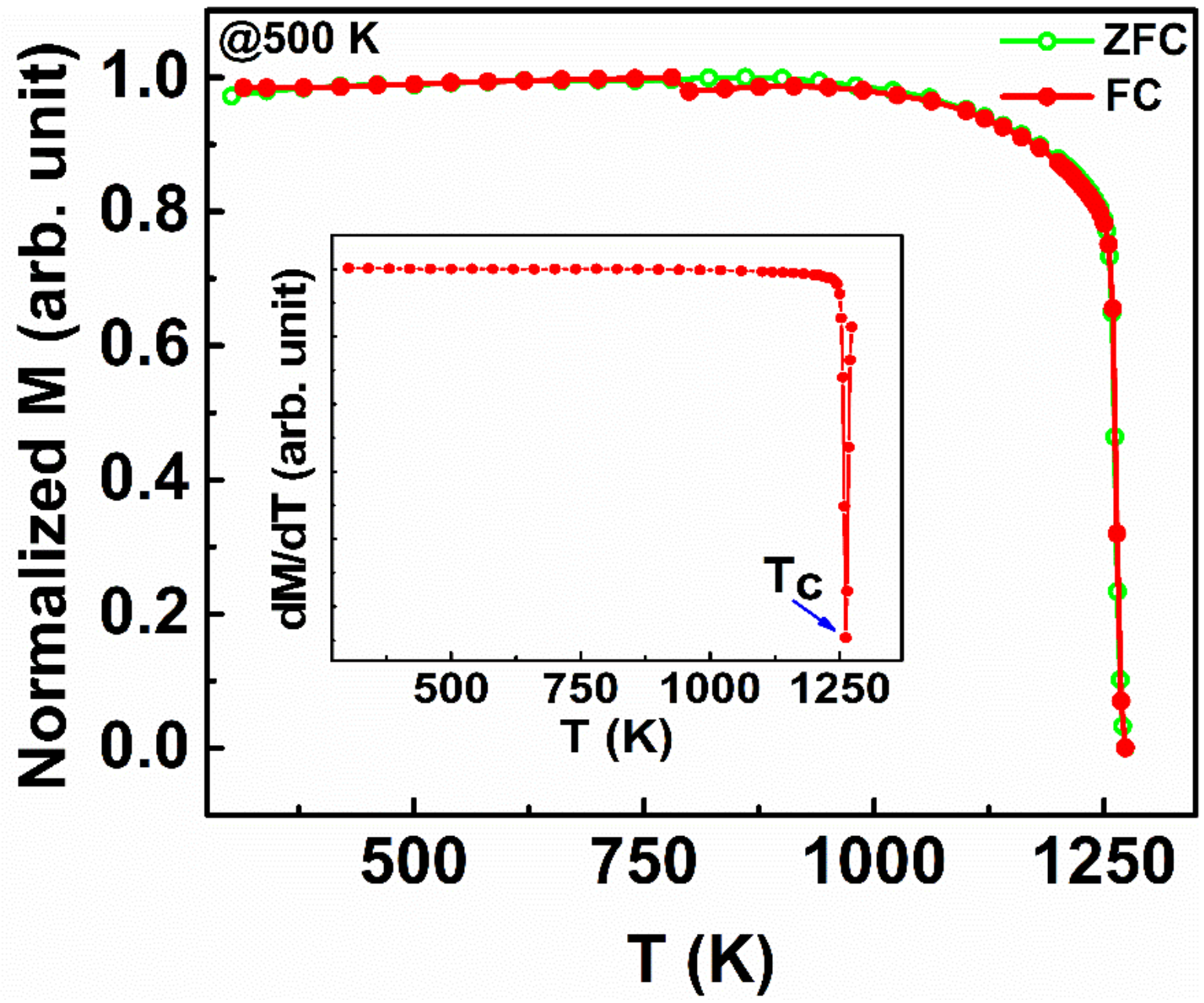

**Figure 3.** Thermo-magnetization curves of the $Co_2FeAl$ nanoparticles in the temperature interval of 300 –1273 K measured at 500 Oe. Inset represents the derivative of M (T) curve of $Co_2FeAl$ for the precise determination of phase transition temperature ($T_c$).

The $T_c$ is calculated from the peak value of dM/dT versus temperature (T) curve, and is found to be 1261 K (see the inset of figure 3). To analyze the MC effect of CFA-NPs, isothermal magnetization M(H) is investigated at various temperatures with a temperature interval of 2 K around the $T_{c,}$ as shown in figure 4(a). The magnetization (M) increases gradually with the field up to 4 kOe and thereafter approaches a constant value. Then the M value decreases with increasing T, suggesting the magnetic transition from a low-T ferromagnetic to a high-T paramagnetic phase. The large change in M at 1251-1253 K isotherms is indicating the possibility of large MC effect of CFA-NPs around this temperature range. In order to see the hysteresis losses and their effects in MCE, we have measured M(H) isotherms in decreasing field (14 kOe -0 Oe) mode (not shown here), and have not found any difference as compared with the M(H) isotherms in increasing field (0 Oe -14 k Oe) mode. The magnetization curves taken during heating and cooling processes near the $T_c$ coincide with each other, exhibiting a negligible hysteresis for all the temperature ranges. Such reversibility of the magnetic transition is essential from the application point of view in magnetic refrigeration.

In order to see whether the Landau mean-field theory of the phase transition is valid for CFA-NPs system, a conventional Arrott plot ($M^2$ vs H/M) [41] is constructed as shown in figure 4(b). The Arrott plot should generate a set of parallel straight lines for the mean field values of the critical exponents $\beta = 0.5$ and $\gamma = 1$. And the line passing through the origin resembles to the isotherm exactly at the critical temperature ($T_c$). As seen from the figure 4(b), Arrott plot does not generate parallel straight lines. Moreover, a substantial non-linearity with downward curvature can be seen even at the high field region. Thus, the mean field theory can't explicate the phase transition of the present system. From Banerjee's criterion [42], the Arrott curves exhibiting a positive slope specify the FM-PM phase transition of second order. The magnetic entropy change, $-\Delta S_M$ is calculated by using Maxwell relation [43]:

$$\Delta S_M = \mathrm{S(T,H)} - \mathrm{S(T,0)} = \int_0^H \left(\frac{\partial M}{\partial T}\right)_H \mathrm{dH} \qquad (1)$$

As the experimental values of magnetization data are only available at the discrete values of the temperatures, the integral in this equation may be replaced by the summation. It can be understood from equation (1) that the magnetic entropy change will be maximum only when ∂M/∂T is maximum. The sign of $-\Delta S_M$ is determined by the sign of ∂M/∂T. When it is positive, it will be called conventional MC effect [44]. The change in magnetic entropy ($-\Delta S_M$) with respect to T is shown in figure 5. The magnetic entropy change show a positive anomaly at 1252 K. A large value of $-\Delta S_M$ ~15 J/Kg-K is detected for change of the magnetic field of 14 kOe.

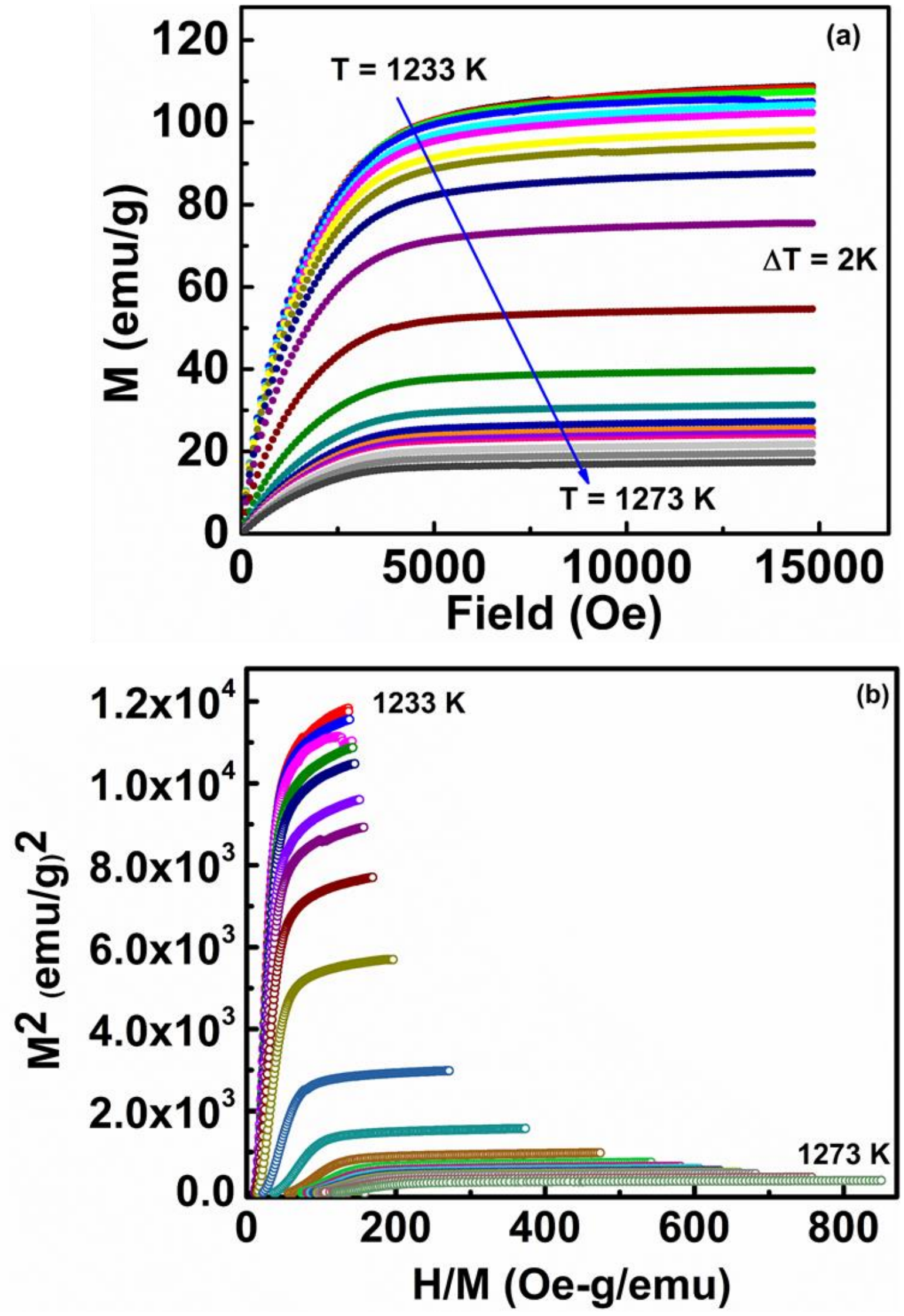

**Figure 4.** (a) Isothermal magnetization M(H) curves of CFA-NPs measured around $T_c$. (b) Arrott plot of the isotherms measured near the phase transition.

The obtained value of $-\Delta S_M$ is comparable in magnitude with the peak value of another GMCE material $Gd_5Si_2Ge_2$ [29], which is very well known in the field of refrigeration technology. It is much larger, i.e., at least thrice the peak value of expensive Gd [29]. Our results show higher or at least comparable MC values as compared to the other Heusler materials such as $Fe_2CoAl$ [45], $Mn_{1-x}Cr_xCoGe$ [46], $MnCo_{1-x}Zr_xGe$ [47] and $Mn_{1-x}Al_xCoGe$ [48]. We also compare our results to the Heusler based thin films (see table 1). As a matter of the fact, the small mass of the thin films makes it challenging to evaluate MC properties quantitatively, particularly the adiabatic temperature change ($\Delta T_{ad}$). Because of this, a limited number of publications are available on MC properties of thin films of metals and compounds. Most of them are concerned with the study of MCE in Ni-Mn-X (X= Ga, In, Sn) thin films. V. Recarte *et al.* [49] deposited NiMnGa film with thickness of 0.4 µm onto alumina substrate. The film exhibited a magnetostructural phase transition at 346 K, and the magnetic entropy change ($-\Delta S_M$) was found to be 8.5 J/Kg-K, at the 60 kOe field. R. Niemann *et al.* [50] observed inverse MC effect in epitaxially grown metamagnetic Ni–Co–Mn–In film (0.2 µm) on MgO (001) substrate. The resulting $-\Delta S_{M,}$ was -8.8 J/kg-K at 353 K for the magnetic field change of 90 kOe. Lately, E. Yüzüak *et al.*[51] detected inverse MCE in epitaxial Ni-Mn-Sn/MgO thin films of 200 nm thickness. The $-\Delta S_M$ and relative cooling power (RCP) were -1.5 J/Kg-K and 33.9 J/Kg at 10 kOe field, respectively. A comparison of all the values, as discussed above, are tabulated in table 1.

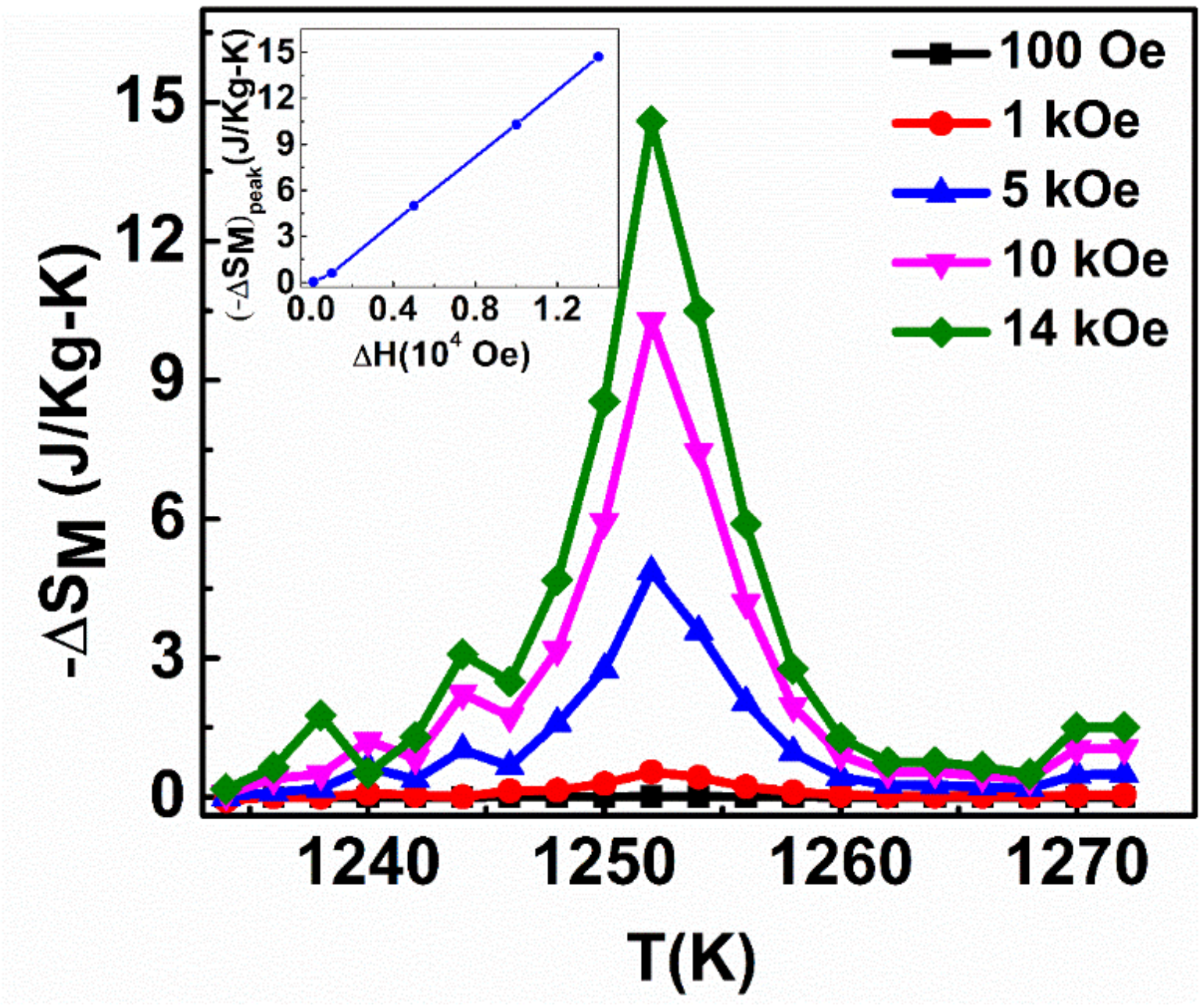

**Figure 5.** Magnetic entropy change ($-\Delta S_M$) with temperature for CFA-NPs; inset figure depicts the change of peak value of $-\Delta S_M$ with respect to the change in field.

As seen from the inset of the figure 5, $-\Delta S_M$ peak height is very sensitive to the field change (ΔH) from H = 5 kOe onwards. It increases linearly relative to ΔH and the width of the peak decreases. RCP is an important parameter to evaluate the usefulness of the material for magnetic refrigeration. This parameter is also used for a comparison with other MC materials. It is a measure of amount of heat transferred between the hot and cold reservoirs in an ideal refrigeration cycle and defined as: RCP = $-\Delta S_{M_{peak}} \times \Delta T_{FWHM}$, where $\Delta T_{FWHM}$ represents full width at half maximum of the $-\Delta S_M$ vs T curve [52]. The large value of RCP of 89 J/Kg (see figure 6) was observed at the change of magnetic field of 14 kOe, which is signifying a large heat conversion capacity in a refrigeration cycle. $\Delta T_{FWHM}$ also represents the span of working-temperature, which is found to be around ~6 K in the present system. As V. I. Zverev et al. [53] and Engelbrecht et al.

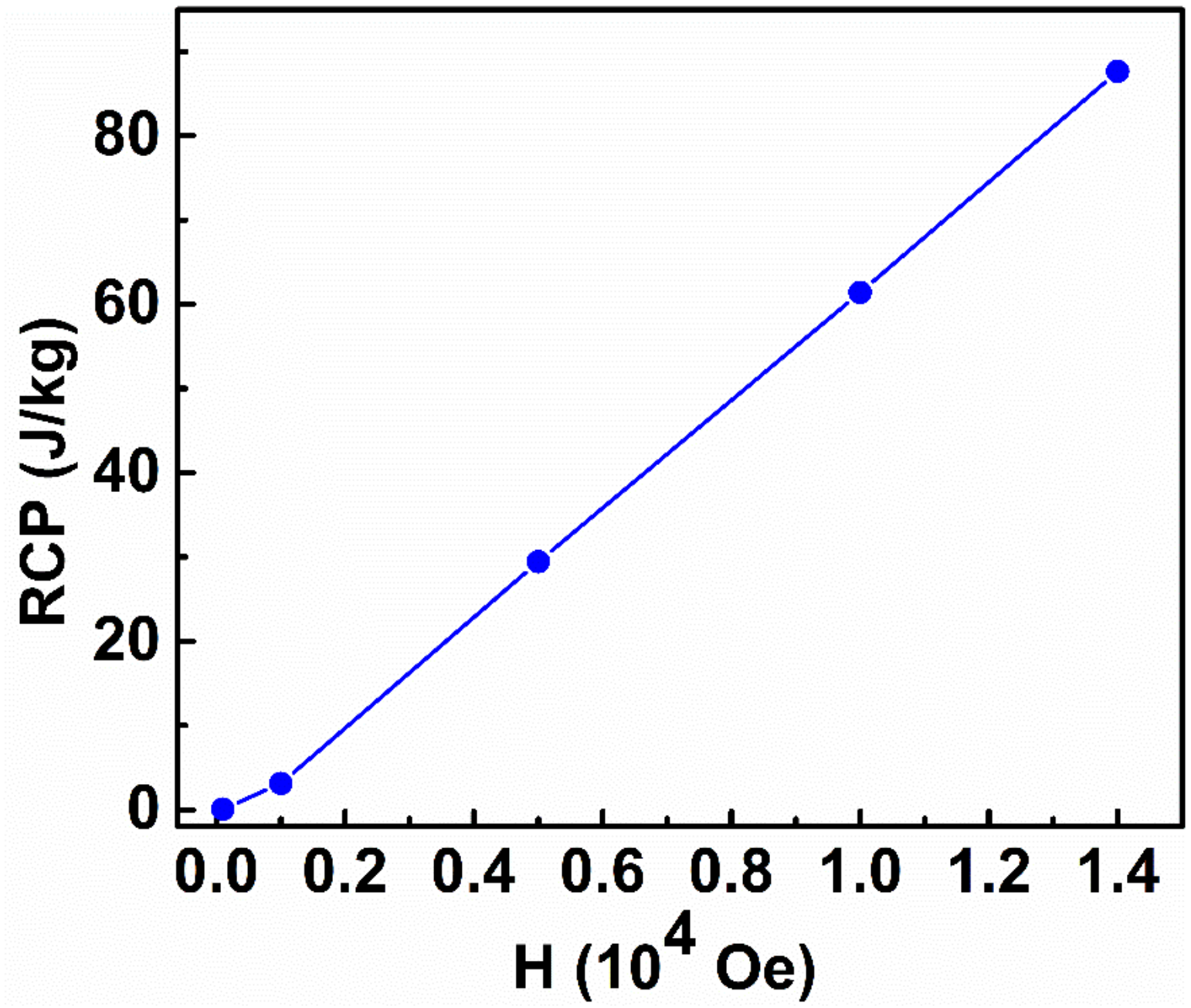

**Figure 6.** Dependence of RCP with magnetic field near the phase transition ($T_c$).

[54] previously suggested that a material having a broad peak of $-\Delta S_M$ is much better than that of materials with a sharp peak of magnetic entropy change for cooling applications. Therefore, in order to fine-tune the $T_c$ and the temperature span of the CFA-NPs, studies of composition optimization are in progress. The refrigeration capacity (RC) and entropy change ($-\Delta S_M$) have shown a linear behavior (see figures 5, 6) with respect to magnetic field. The increasing nature of the peak value of the entropy change with ΔH can be suitable for Ericsson-cycle refrigeration application [55]. And therefore the study of magnetic and MC properties on $Co_2$-based Heusler alloys might be useful for the

**Table 1.** A comparison of the magnetic entropy change ($-\Delta S_M$), RCP, ΔH, $T_c$, and nature of the phase transition with other magnetocaloric materials.

| **Materials** | | **$-\Delta S_M$ (J/Kg-K)** | **RCP (J/Kg)** | **Field range (ΔH) in kOe** | **$T_c$ (K)** | **Nature of the Phase transition** | **Refs.** |
|---|---|---|---|---|---|---|---|
| | $Co_2FeAl$ | ~15 | 89 | 14 | 1261 | second-order | present work |
| **Heusler alloy nanoparticles** | $Fe_2CoAl$ | 2.65 | 44 | 20 | 830 | second-order | [45] |
| **Bulk Heusler alloys** | $Mn_{1-x}Cr_xCoGe$ | ~28.5 | - | 50 | 322 | first-order | [46] |
| | $MnCo_{1-x}Zr_xGe$ | 7.2 | 266 | 50 | 274 | first-order | [47] |
| | $Co_2Cr_{0.25}Mn_{0.75}Al$ | 3.8 | ~285 | 90 | 720 | second order | [13] |
| | $Mn_{1-x}Al_xCoGe$ | 12 | 303 | 50 | 286 | first-order | [48] |
| **Giant MC Materials** | $Gd_5Si_2Ge_2$ | 18 | - | 20 | 276 | first order | [29] |
| | $(MnNiSi)_{1-x}(FeCoGa)_x$ | 25 | 191.8 | 50 | 323 | first-order | [16] |
| | Fe-Rh | 20 | - | 20 | | first-order | [30] |
| **MCE in Heusler thin films** | NiMnGa | 8.5 | - | 60 | 346 | - | [49] |
| | Ni-Co-Mn-In | -8.8 | - | 90 | 353 | - | [50] |
| | Ni-Mn-Sn | -1.5 | 33.9 | 10 | ~ 270 | - | [51] |
| **Others** | $Fe_{68.5}Mo_5Si_{13.5}B_9Cu_1Nb_3$ | ~1.1 | 63 | 15 | ~ 475 | - | [27] |
| | $Fe_{50}Rh_{50}$ particles | 9.7 | 230 | 30 | - | first-order | [56] |

application of multistage magnetic refrigeration [13]. From a simple linear extrapolation of our data to 50 k Oe, will give an approximate value of RCP = 305 J/Kg, and $-\Delta S_M$ = 50 J/Kg-K. On a comparison of recently reported high-T MC material $Co_2Cr_{1-x}Mn_xAl$ Heusler alloy [13], having $T_c$ of 720 K on the same field scale (i.e. 50 kOe), our values of entropy change ($-\Delta S_M$) is 20 times larger and RC value is slightly larger or at least comparable with it.

Franco et al. [57] proposed a master curve that is used to see the nature of the magnetic phase transition. This curve can be plotted as the normalized entropy change ($\Delta S_M^{'}$) where $\Delta S_M^{'} = \Delta S_M / \Delta S_M^{peak}$, with respect to the rescaled temperature (θ) defined by

$$\theta = \left\{ \frac{-(T-T_c)}{(T_{r1}-T_c)} \text{ , for } T \leq T_C; \ \frac{(T-T_c)}{(T_{r2}-T_c)} \text{ , for } T > T_C \right\} \qquad (2)$$

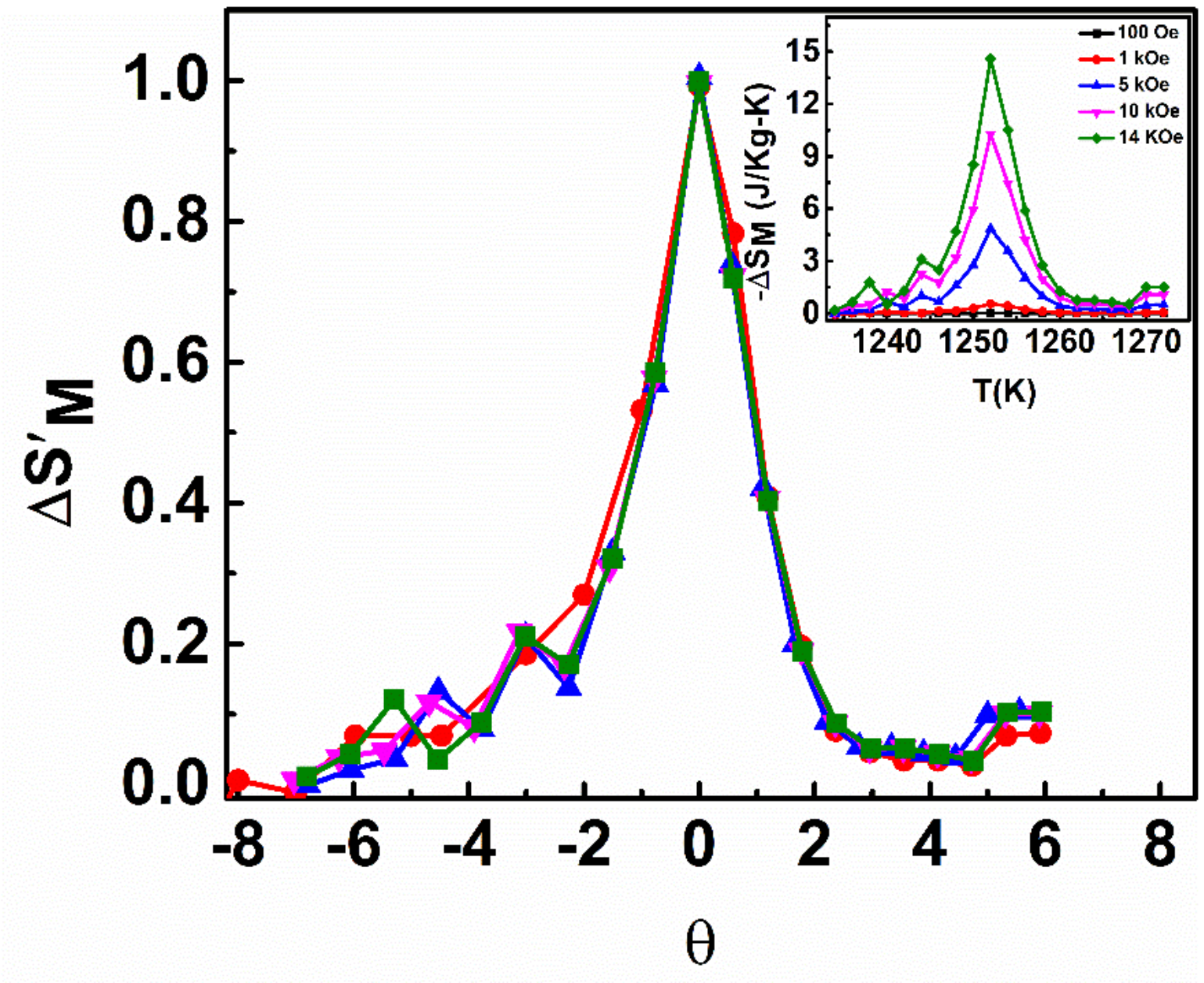

**Figure 7.** The normalized entropy change ($\Delta S_M^{'}$) as a function of rescaled temperature (θ) near the phase transition for CFA-NPs; inset shows $-\Delta S_{M,}$ versus T at different fields.

In eqn. 2, $T_c$ represents the temperature at which $\Delta S_M^{peak}$ is attained. The reference temperatures $T_{r1}$ and $T_{r2}$ below and above the $T_c$, are taken as the temperatures corresponding to $\frac{1}{2}\Delta S_M^{peak}$. In figure 7, all the curves ($\Delta S_M^{'}$ vs θ) of CFA-NPs, taken at different magnetic fields, are merged into a single universal curve (see the inset of figure 7 for a comparison), which confirms a second-order magnetic phase transition. Such type of phenomenological curves was previously reported for the materials exhibiting second-order phase transition [57-59]. The present study suggests a possibility of using this material in high temperature cooling applications. As in multi-stage magnetic refrigeration, cooling from high temperature is desired in more than one stage [60]. Such multistage refrigerators are highly suitable in industry, where the low temperature of one stage acts as upper temperature for the next stage [60, 61].

## 4. Conclusions

We have performed a detailed investigation on magnetic phase transition and magnetocaloric effect in $Co_2FeAl$ nanoparticles. The positive slope of the Arrot curve suggests the second order FM to PM phase transition near the $T_c$. And the normalized entropy change ($\Delta S_M^{'}$) with respect to the rescaled temperature (θ) at different fields collapsed into a single universal curve, which further confirms the phase transition to be of second-order. Owing to abrupt change in magnetization, it is conclusive that CFA-NPs will exhibit GMCE across the $T_c$. The observed peak value is superior or at least comparable with that of the other MC materials on a comparison of the peak heights. The present study provides a window for future research on tuning the transition temperature ($T_c$) and MC effect of $Co_2$-based Heusler nanoalloys by controlling the elemental composition and/or partial substitution of other elements.

**Data availability statement**

The data that support the findings of this study are available upon reasonable request from the corresponding authors.

**Acknowledgements:** A. Ahmad acknowledges University Grants Commission, New Delhi and Ministry of Education (MoE), India for providing the research fellowship. Amal Kumar Das acknowledges the financial support from Department of Science and Technology (DST), India (project no. EMR/2014/001026). The fruitful discussion with Dr. Ramchandra Sahoo is highly acknowledged.

## References

[1] Z. Wen, H. Sukegawa, S. Kasai, K. Inomata, and S. Mitani 2014 Tunnel magnetoresistance and spin-transfer-torque switching in polycrystalline $Co_2FeAl$ full-Heusler alloy magnetic tunnel junctions on amorphous $Si/SiO_2$ substrates *Phys. Rev. Appl.* **2** 024009.

[2] I. Galanakis, P. Dederichs, and N. Papanikolaou 2002 Slater-Pauling behavior and origin of the half-metallicity of the full-Heusler alloys, *Phys. Rev. B* **66** 174429.

[3] A. Ahmad, S.K. Srivastava and A.K. Das 2019 Effect of $L2_1$ and XA ordering on phase stability, half-metallicity and magnetism of $Co_2FeAl$ Heusler alloy: GGA and GGA+ U approach *J. Magn.Magn. Mater*. **491** 165635.

[4] A. Ahmad, A.K. Das, and S.K. Srivastava, 2020 Competition of $L2_1$ and XA ordering in $Fe_2CoAl$ Heusler alloy: a first-principles study, *The Eur. Phys. J. B* **93** pp 1-7.

[5] A. Ahmad, S.K. Srivastava and A.K. Das 2020 Phase stability and the effect of lattice distortions on electronic properties and half-metallic ferromagnetism of $Co_2FeAl$ Heusler alloy: An ab initio study, J. Phys.: Condens. Matter **32** 415606.

[6] A. Ahmad, S. Guchhait, H. Ahmad, S.K. Srivastava, and A.K. Das 2019 First-principles investigation of the $L2_1$ and XA ordering competition in $Co_2FeAl$ Heusler alloy *AIP Confer. Proc.* **2142** 090006.

[7] A. Ahmad, S.K. Srivastava and A.K. Das 2021 First-principles calculations and experimental studies on $Co_2FeGe$ Heusler alloy nanoparticles for spintronics applications *J. Alloys Compd.* **878** 160341.

[8] K. Kilian, and R.H. Victora 2000 Electronic structure of $Ni_2MnIn$ for use in spin injection *J. Appl. Physics* **87** 7064-7066.

[9] S. Datta, B. Das, 1990 Electronic analog of the electro-optic modulator *Appl. Phys. Lett*. **56** pp 665-667.

[10] R. Kainuma, Y. Imano, W. Ito, Y. Sutou, H. Morito, S. Okamoto, O. Kitakami, K. Oikawa, A. Fujita and T. Kanomata 2006 Magnetic-field-induced shape recovery by reverse phase transformation, Nat. **439** pp 957-960.

[11] P. Devi, M.G. Zavareh, C.S. Mejía, K. Hofmann, B. Albert, C. Felser, M. Nicklas and S. Singh 2018 Reversible adiabatic temperature change in the shape memory Heusler alloy $Ni_{2.2}Mn_{0.8}Ga$ An effect of structural compatibility *Phys. Rev. Mater.* **2** 122401.

[12] T. Graf, J. Barth, B. Balke, S. Populoh, A. Weidenkaff and C. Felser 2010 Tuning the carrier concentration for thermoelectrical application in the quaternary Heusler compound $Co_2TiAl_{(1-x)}Si_x$ *Scripta Materialia* **63** pp 925-928.

[13] P. Nehla, V. Anand, B. Klemke, B. Lake, R. Dhaka 2019 Magnetocaloric properties and critical behavior of $Co_2Cr_{1-x}Mn_xAl$ Heusler alloys *J. Applied Physics* **126** 203903.

[14] J. Waybright, L. Halbritter, B. Dahal, H. Qian, Y. Huh, P. Lukashev, and P. Kharel 2019 Structure and magnetism of $NiFeMnGa_xSn_{1-x}$ (x= 0, 0.25, 0.5, 0.75, 1.00) Heusler compounds, *AIP Advanc.* **9** 035105.

[15] Y. Jin, J. Waybright, P. Kharel, I. Tutic, J. Herran, P. Lukashev, S. Valloppilly and D. Sellmyer 2017 Effect of Fe substitution on the structural, magnetic and electron-transport properties of half-metallic $Co_2TiSi$ *AIP Adv.* **7** 055812.

[16] S. Ghosh, A. Ghosh, P. Sen, K. Mandal 2020 Giant Room-Temperature Magnetocaloric Effect Across the Magnetostructural Transition in $(Mn\ Ni\ Si)_{1-x}(Fe\ Co\ Ga)_x$ Alloys *Phys. Rev. Appl.* **14** 014016.

[17] R. De Groot, F. Mueller, P. Van Engen and K. Buschow 1983 New class of materials: half-metallic ferromagnets, *Phys. Rev. Lett.* **50** 2024.

[18] T. Graf, C. Felser, and S.S. Parkin 2011 Simple rules for the understanding of Heusler compounds *Prog. Solid State Chem.* **39** (2011) pp 1-50.

[19] Y. Srivastava, S. Rathod, P.K. Singh, S.K. Vajpai, and S. Srivastava 2018 Study of magneto-structural phase transitions and magnetocaloric effects in Co-based Heusler alloys synthesized via mechanical milling *J. Magn. and Magn. Mater.* **462** pp 195-204.

[20] A. Ahmad, S. Mitra, S.K. Srivastava, A.K. Das 2019 Size-dependent structural and magnetic properties of disordered $Co_2FeAl$ Heusler alloy nanoparticles, *J. Magn. and Magn. Mater.* **474** pp 599-604.

[21] S. Wurmehl, G.H. Fecher, H.C. Kandpal, V. Ksenofontov, C. Felser and H.J. Lin, Investigation of $Co_2Fe$ Si: The Heusler compound with highest Curie temperature and magnetic moment 2006 *Appl. Phys. Lett.* **88** 032503.

[22] J. Kübler, G. Fecher and C. Felser 2007 Understanding the trend in the Curie temperatures of $Co_2$-based Heusler compounds: Ab initio calculations *Phys. Rev. B* **76** 024414.

[23] X. Zhang, H. Zhang, M. Qian and L. Geng 2018 Enhanced magnetocaloric effect in Ni-Mn-Sn-Co alloys with two successive magnetostructural transformations, *Sci. Rep*. **8** pp 1-11.

[24] T. Krenke, E. Duman, M. Acet, E.F. Wassermann, X. Moya, L. Mañosa, A. Planes, Inverse magnetocaloric effect in ferromagnetic Ni–Mn–Sn alloys, Nature materials, 4 (2005) 450-454.

[25] L. Huang, D. Cong, L. Ma, Z. Nie, Z. Wang, H. Suo, Y. Ren and Y. Wang 2016 Large reversible magnetocaloric effect in a Ni-Co-Mn-In magnetic shape memory alloy *Appl. Phys. Lett.* **108** 032405.

[26] K. Gschneidner Jr and V.K. Pecharsky 2020 Magnetocaloric materials, *An. Rev. Mater. Scie.* **30** pp 387-429.

[27] V. Franco, J. Blázquez, C. Conde and A. Conde 2006 A Finemet-type alloy as a low-cost candidate for high-temperature magnetic refrigeration, *Appl. Phys. Lett*. **88** 042505.

[28] C. Zhang, D. Wang, Q. Cao, Z. Han, H. Xuan and Y. Du 2008 Magnetostructural phase transition and magnetocaloric effect in off-stoichiometric $Mn_{1.9-x}Ni_xGe$ alloys *Appl. Phys. Lett*. **93** 122505.

[29] V.K. Pecharsky and K.A. Gschneidner Jr. 1997 Giant magnetocaloric effect in $Gd_5(Si_2Ge_2)$, *Phys. Rev. Lett*. **78** 4494.

[30] R. Gimaev, A. Vaulin, A. Gubkin, V. Zverev 2020 Peculiarities of Magnetic and Magnetocaloric Properties of Fe–Rh Alloys in the Range of Antiferromagnet–Ferromagnet Transition, Phys. Metals Metallography **121** pp 823-850.

[31] J.D. Moore, K. Morrison, G.K. Perkins, D.L. Schlagel, T.A. Lograsso, K.A. Gschneidner, V.K. Pecharsky and L.F. Cohen 2009 Metamagnetism Seeded by Nanostructural Features of Single-Crystalline $Gd_5Si_2Ge_2$ *Adv. Mater*. **21** pp 3780-3783.

[32] A. Smith, C.R. Bahl, R. Bjørk, K. Engelbrecht, K.K. Nielsen and N. Pryds 2012 Materials challenges for high performance magnetocaloric refrigeration devices *Adv. En. Mater*. **2** pp 1288-1318.

[33] V. Basso, C.P. Sasso, K.P. Skokov, O. Gutfleisch and V.V. Khovaylo 2012 Hysteresis and magnetocaloric effect at the magnetostructural phase transition of Ni-Mn-Ga and Ni-Mn-Co-Sn Heusler alloys *Phys. Rev. B* **85** 014430.

[34] R. Sahoo, A.K. Nayak, K. Suresh and A. Nigam 2011 Effect of Si and Ga substitutions on the magnetocaloric properties of NiCoMnSb quaternary Heusler alloys *J. Appl. Phys.* **109** 07A921.

[35] J. Panda, S. Saha and T. Nath 2015 Critical behavior and magnetocaloric effect in $Co_{50-x}Ni_xCr_{25}Al_{25}$ (x= 0 and 5) full Heusler alloy system, *J. Alloys Compd*. **644** pp 930-938.

[36] S. Datta, S. Guha, S.K. Panda and M. Kar 2020 Correlation between Critical Behavior and Magnetocaloric Effect near Paramagnetic to Ferromagnetic Phase Transition of $Co_2TiAl_{0.75}Si_{0.25}$ Heusler Alloy *Physica Status Solidi b* **257** 2000123.

[37] N. Telling, P.S. Keatley, G. van der Laan, R. Hicken, E. Arenholz, Y. Sakuraba, M. Oogane, Y. Ando, K. Takanashi and A. Sakuma 2008 Evidence of local moment formation in Co-based Heusler alloys, *Phys. Rev. B* **78** 184438.

[38] C. Wang, Y. Guo, F. Casper, B. Balke, G. Fecher, C. Felser, Y. Hwu 2010 Size correlated long and short range order of ternary $Co_2FeGa$ Heusler nanoparticles *Appl. Phys. Lett.* **97** 103106.

[39] H. Elmers, S. Wurmehl, G. Fecher, G. Jakob, C. Felser and G. Schönhense 2004 Field dependence of orbital magnetic moments in the Heusler compounds $Co_2FeAl$ and $Co_2Cr_{0.6}Fe_{0.4}Al$ *Appl. Phys. A* **79** pp 557-563.

[40] S. Roy, N. Khan, R. Singha, A. Pariari and P. Mandal 2019 Complex exchange mechanism driven ferromagnetism in half-metallic Heusler $Co_2TiGe$: Evidence from critical behavior, *Phys. Rev. B* **99** 214414.

[41] A. Arrott 1957 Criterion for ferromagnetism from observations of magnetic isotherms Phys. Rev. **108** 1394.

[42] B. Banerjee 1964 On a generalised approach to first and second order magnetic transitions, *PhL* **12** pp 16-17.

[43] T. Krenke, E. Duman, M. Acet, E.F. Wassermann, X. Moya, L. Manosa and A. Planes 2005 Inverse magnetocaloric effect in ferromagnetic Ni-Mn-Sn alloys *Nat. Mater*. **4** 450-454.

[44] T. Chabri, A. Ghosh, S. Nair, A. Awasthi, A. Venimadhav and T. Nath, 2018 Effects of the thermal and magnetic paths on first order martensite transition of disordered $Ni_{45}Mn_{44}Sn_9In_2$ Heusler alloy exhibiting a giant magnetocaloric effect and magnetoresistance near room temperature *J. Phys. D: Appl. Phys*. **51** 195001.

[45] A. Ahmad, S. Mitra, S.K. Srivastava and A.K. Das 2021 Structural, magnetic, and magnetocaloric properties of intermetallic $Fe_2CoAl$ Heusler nanoalloy *arXiv:2102.11195*.

[46] N. Trung, V. Biharie, L. Zhang, L. Caron, K. Buschow and E. Brück 2010 From single-to double-first-order magnetic phase transition in magnetocaloric $Mn_{1-x}Cr_xCoGe$ compounds, *Appl. Phys. Lett.* **96** 162507.

[47] A. Aryal, A. Quetz, S. Pandey, I. Dubenko, S. Stadler and N. Ali 2017 Phase transitions and magnetocaloric properties in $MnCo_{1-x}Zr_xGe$ compounds *Adv. Cond. Matter Phys*. 2017 pp 1-6.

[48] A. Aryal, A. Quetz, S. Pandey, T. Samanta, I. Dubenko, M. Hill, D. Mazumdar, S. Stadler, N. Ali 2017 Magnetostructural phase transitions and magnetocaloric effects in as-cast $Mn_{1-x}Al_xCoGe$ compounds *J. Alloys Compd.* **709** pp 142-146.

[49] V. Recarte, J. Pérez-Landazábal, V. Sánchez-Alárcos, V. Chernenko, M. Ohtsuka 2009 Magnetocaloric effect linked to the martensitic transformation in sputter-deposited Ni–Mn–Ga thin films *Appl. Phys. Lett*. **95** 141908.

[50] R. Niemann, O. Heczko, L. Schultz and S. Fähler 2010 Metamagnetic transitions and magnetocaloric effect in epitaxial Ni–Co–Mn–In films *Appl. Phys. Lett.* **97** 222507.

[51] E. Yüzüak, I. Dincer, Y. Elerman, A. Auge, N. Teichert and A. Hütten 2013 Inverse magnetocaloric effect of epitaxial Ni-Mn-Sn thin films *Appl. Phys. Lett*. **103** 222403.

[52] A. Aryal, A. Quetz, S. Pandey, T. Samanta, I. Dubenko, D. Mazumdar, S. Stadlera and N. Ali 2016 Phase transitions and magnetocaloric and transport properties in off-stoichiometric $GdNi_2Mnx$, *J. Appl. Phys*. **119** 043905.

[53] V. Zverev, A. Tishin, M. Kuz'Min 2010 The maximum possible magnetocaloric Δ T effect *J. Appl. Phys.* **107** 043907.

[54] K. Engelbrecht and C.R.H. Bahl 2010 Evaluating the effect of magnetocaloric properties on magnetic refrigeration performance *J. Appl. Phys*. **108** 123918.

[55] H. Takeya, V. Pecharsky, K. Gschneidner Jr, J. Moorman 1994 New type of magnetocaloric effect: Implications on low-temperature magnetic refrigeration using an Ericsson cycle *Appl. Phys. Lett.* **64** pp 2739-2741.

[56] Y. Cao, Y. Yuan, Y. Shang, V.I. Zverev, R.R. Gimaev, R. Barua, R. Hadimani, L. Mei, G. Guo, H. Fu 2020 Phase transition and magnetocaloric effect in particulate Fe-Rh alloys *J. Mater. Sci.* **55** pp 13363-13371.

[57] V. Franco, J. Blázquez, A. Conde 2006 Field dependence of the magnetocaloric effect in materials with a second order phase transition: A master curve for the magnetic entropy change *Appl. Phys. Lett*. **89** 222512.

[58] B. Dahal, C. Huber, W. Zhang, S. Valloppilly, Y. Huh, P. Kharel, D. Sellmyer 2019 Effect of partial substitution of In with Mn on the structural, magnetic, and magnetocaloric properties of $Ni_2Mn_{1+x}In_{1-x}$ Heusler alloys *J. Phys. D: Appl. Phys*. **52** 425305.

[59] V. Franco, A. Conde, J. Romero-Enrique, Y. Spichkin, V. Zverev, A. Tishin 2009 Field dependence of the adiabatic temperature change in second order phase transition materials: Application to Gd *J. Appl. Phys.* **106** 103911.

[60] A. Kitanovski and P.W. Egolf 2010 Innovative ideas for future research on magnetocaloric technologies, Intern. J. Refrigeration **33** pp 449-464.

[61] V. Franco, J. Blázquez, J. Ipus, J. Law, L. Moreno-Ramírez and A. Conde 2018 Magnetocaloric effect: From materials research to refrigeration devices *Prog. Mater. Sci.* **93** pp 112-232.